\documentclass[aps,pra,twocolumn,superscriptaddress,noshowpacs,hyperref]{revtex4-1}
\usepackage{amssymb}
\usepackage{amsmath}
\usepackage{amsfonts}
\usepackage{bm}
\usepackage{graphicx}
\usepackage[dvipsnames]{xcolor}
\usepackage{calc} 
\usepackage{epsfig}
\usepackage{epstopdf}
\usepackage{natbib}
\begin{document}





\title{The Effective Model of the Molecule Graphene System and One Application beyond Graphene}

\author{Shuai Li}

\author{Wen-Xuan Qiu}

\author{Jin-Hua Gao}
\affiliation{School of Physics and Wuhan National High Magnetic field center,
Huazhong University of Science and Technology, Wuhan 430074,  China}
\email{jinhua@hust.edu.cn}

\begin{abstract}
Recently, a new kind of two dimensional (2D) artificial electron lattice, i.e. molecule graphene, has drawn a lots of interest, where the metal surface electrons are transformed into a honeycomb lattice via absorbing a molecule lattice on metal surface\cite{mg2012, lin2014}. In this work, we would like to point out that the technique used to build molecule graphene actually gives a promising way to explore the interesting physics of other novel 2D lattice beyond graphene.  The key issue is that this system is an antidot system, where the absorbed molecule normally gives  a repulsive potential. So, we need to establish a map between the molecule arrangement and the desired surface electron lattice. To give a concrete example, we first analyse the effective models of molecule graphene, and estimate the model parameters for the Cu/CO system through  numerical fitting the exerimental data. Then, we design a molecule lattice, and show that, with this kind of lattice, the surface electrons can be transformed into a Kagome like lattice. Using the estimated parameters of Cu/CO system, we calculate the corresponding energy bands and LDOS, which can be readily tested in experiment.  We hope that our work can  stimulate further theoretical and experimental interest in this novel artificial 2D electron lattice.
\end{abstract}
\maketitle

\section{Introduction}
Recently, a novel kind of artificial two dimensional (2D) electronic lattice, named molecule graphene, has been successfully realized on metal surface  in order to mimic the Dirac like linear dispersion of  graphene\cite{mg2012,lin2014}.  In experiment\cite{mg2012},  CO molecules are absorbed on Cu(111) surface and then  assembled into a hexagonal lattice by the tip of scanning tunneling microscopy (STM). The $\textrm{Cu}(111)$ surface  has a Shockley surface state near the Fermi level, which can be viewed as a two dimensional electron gas with $k^2$ dispersion. Meanwhile, each absorbed molecule applies a local potential on the surface states, and thus the molecule lattice is equivalent to a lateral periodic potential, which transforms the surface electrons into an artificial honeycomb lattice\cite{louie2016}.

Artificial 2D lattice systems are always the research focus during the  last two decades,  because that it not only relates to the fundamental physics about 2D system, but also has potential for application.  The nano-patterned semiconductor 2DEG, cold atom, and the photonic crystal are the well-known artificial 2D lattice systems, which have been intensively studied. Here, the metal surface state patterned by the absorbed molecules offers a new artificial 2D lattice system, where molecule graphene is first example realized in experiment.  This new artificial 2D system has its own characteristic and merits.  First, it is a 2D electron lattice in solid state system in contrast to the cold atom and photonic crystal systems, so that the electron-electron interaction can be included naturally.  Moreover, because the 2D electrons are the surface states, the STM gives a powerful detecting method, which not only can measure the DOS, LDOS, but also could get the accurate information about the edge states\cite{lin2014},  charge order, the spin order (by spin-polarized STM\cite{sp-stm}), and even the quasi particle energy (by inelastic electron tunnelling spectroscopy).  These critical information can not be easily detected in other artificial 2D lattice systems.
Considering these great advantages, this new artificial 2D lattice system deserves more research efforts.

So far, all the investigations about the molecule graphene system (i.e. metal surface electrons patterned by absorbed molecule lattice) are mainly focused on the physics of graphene, e.g. the linear dispersion\cite{mg2012,DFT2014,scattering2013,anisotropic2015}, edge and impurity states\cite{lin2014}, finite size effect\citep{finitesize2014},  topological nontrivial phases\cite{nonabelian2013,qsh2012} and superconductivity\cite{SC2014}.
However,  it should be noticed  that this powerful technique actually provides  a promising method to study interesting physics in any 2D electron lattice, not just the graphene lattice. To the best of our knowledge, there are still few works to discuss the possibility to construct other novel 2D lattices on metal surface by this technique.  One critical issue is that, like the Cu/CO system,  the absorbed molecule normally applies a repulsive potential on the metal surface state, so that it is an antidot system. To get a desired 2D lattice of surface electrons, we need to design a special molecule arrangement, and thus numerical simulation with realistic parameters  becomes indispensable.

In this work,   we first theoretically analyse the two effective models of the molecule graphene used in literatures, discuss their relation and differences.  By numerical simulation and comparing with the experimental data, we estimate the parameters of different effective models for the Cu/CO system, which is an ideal system to construct 2D lattice of surface electron.  We point out that the potential value of  the absorbed CO molecules in the muffin-tin model is of the order of several eV, which is a critical parameter to simulate other 2D lattice in Cu/CO system.
Then, with these estimated parameters, we give an interesting example which can be readily tested in experiment. We design a 2D molecule lattice  and show that, with this kind of molecule arrangement, we can get a band structure rather similar to that of Kagome lattice, which is of special interest in condensed matter physics. We hope that our work can stimulate further theoretical and experimental interests in this promising artificial 2D electronic lattice system.

The paper is organized as following: In Sec. II, we discuss the two effective models of molecule graphene, and estimate the parameters of the effective models  for the Cu/CO system; In Sec. III, we design a molecule arrangement and show that it can give a Kagome like band structure via numerical simulation; Finally, we give a summary in Sec. IV.

\section{Effective models of molecule graphene}
Basically, the molecule graphene is one kind of artificial graphene, which was first proposed in the semiconductor 2DEG system\citep{louie2016,2deg2009,polini2013}. In the literatures, except the DFT calculation\cite{DFT2014},  two effective models using different potential models are used to describe   such artificial graphene system. For the artificial graphene, they both work and are equivalent in some sense. However, only the effective model with muffin-tin potential is a general model which is applicable to all the 2D  lattice. In addition, we  emphasize that the potential values of the two effective models are of different physical meanings, and of different order of magnitude. Sometimes, they are confused in literatures since which effective model is used is not declared.

In order to do a realistic simulation for other 2D lattices, we first need to get some critical parameters of the Cu/CO system by analysing the effective models and the experimental results of the molecule graphene. Here, we only focus on the Cu/CO system, since it has been realized in experiment and is an ideal system to build other 2D lattice. For other systems, e.g. other molecules\cite{lin2014}, simulation can be done in the same way.

\subsection{Two Potential Models}
\begin{figure}
\centering
\includegraphics[width=8cm]{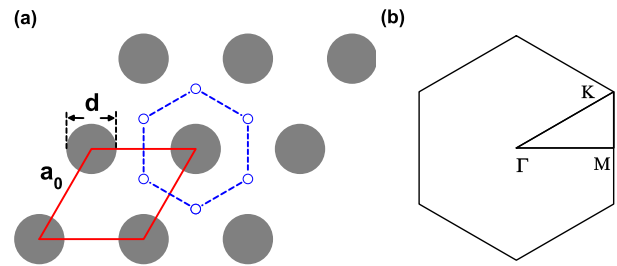}
\caption{(Color online)(a) A muffin-tin potential with hexagonal arrangement. Potential $U_0>0$ inside the gray disks, and zero elsewhere. Blue circles indicate the sites of electrons in the thin film of the surface. Red lines illustrate the unit cell of the molecule lattice. Lattice constant $a_0$, the distance from the center of one molecule to another nearest. $d$, the diameter of the molecule potential disk. (b) The corresponding Brillouin zone of the molecule lattice.
}
\label{muffintin}
\end{figure}

To describe the molecule graphene system, the Cu(111) surface state  is approximately considered as a 2DEG. When the absorbed molecule lattice is formed, it exerts a lateral periodic potential on the surface electrons.  So far, in the literatures, there are  two kinds of potential models used to simulate the lateral periodic potential. One is the muffin-tin potential model\cite{louie2016,lin2014},  as shown in Fig. \ref{muffintin}.   The Hamiltonian of the whole system is
\begin{equation}
H=\frac{\hbar^2}{2m^*}k^2 + U(r),
\end{equation}
where the muffin-tin potential $U(r)$ is  $\textrm{U}_0>0$ inside the gray disks and zero elsewhere, $m^*=0.38m_0$  is the effective mass of the Cu(111) surface state, and the Fermi energy relative to the band bottom is $E_f = 450$ meV.  To calculate the energy bands, we can expend the Hamiltonian with the plane wave basis and get the central equations
\begin{equation}
[\frac{\hbar^2}{2m^*} (k-G)^2 - \epsilon]c_{k-G} + \sum_{G'} U_{G'-G} \cdot c_{k-G'} =0.
\end{equation}
 $G$ and $G'$ are the reciprocal lattice in momentum space, $c_{k}$ is the coefficient of the plane wave and $U_{G}$ is the Fourier component of the muffin-tin potential. As shown in Fig. \ref{muffintin} (a), there are three parameters for the muffin-tin potential: the  lattice constant of the molecule lattice $a_0$, the diameter of the potential disk $d$ and the potential value $\textrm{U}_0$. We can get an analytical expression about the Fourier component of the muffin-tin potential
\begin{equation}\label{ug}
U_{G}=\frac{\pi d}{\Omega |G|} J_1 (|G|\frac{d}{2})\textrm{U}_0 .
\end{equation}
Here, $\Omega$ is the unit cell area of the molecule lattice, $J_1$ is the Bessel function. Given the muffin-tin potential, we can get the energy bands of the molecule graphene by solving the central equations above. The Dirac like linear dispersion is only related with the lowest three bands.

The other potential model for the molecule graphene  is actually a simplification of the muffin-tin potential.  When calculating the energy bands with the central equations,
it is shown that, for the molecule graphene case,   the first Fourier component $U_{G_0}$  dominates, where $|G_0| = |b_1|=|b_2|=|b_1-b_2|$, $b_1$ and $b_2$ are the basis of reciprocal lattice. Thus, in some theoretical literatures, only the $U_{G_0}$ terms are considered as the potential model to simulate the molecule graphene, which we call the $U_{G_0}$ model here\cite{mg2012,polini2013,njp2012}.
The benefit of the $U_{G_0}$ model is that there is now only one potential parameter, i.e. the value of $U_{G_0}$, for the band calculation instead of the two  in the muffin-tin potential case ($U_0$ and $d$).  Note that the molecule distance $a_0$ is the same in any potential models.  So, it is convenient to use the $U_{G_0}$ model  for    both the numerical calculation and
 the fitting of the experiment. We want to emphasize here that when we talk about the potential value of molecule graphene system, it should be aware of which potential model is used. Otherwise, it is easier to confuse $U_0$ and $U_{G_0}$, since they all have the dimension of energy and are both named as potential value.

The disadvantage of the $U_{G_0}$ model is that it is not transferable for studying other 2D lattice, which is one of the main purposes of this work. On one side, there is no direct relation between the $U_{G_0}$ terms of different 2D lattices, even if the same molecules are used. On the other side, for some 2D lattices, only including the $U_{G_0}$ terms is not enough to correctly describe the energy bands.  It should be noticed that the muffin-tin potential model is naturally applicable for all kinds of 2D lattice, though it is not as convenient as the $U_{G_0}$ model for the experiment fitting since more  potential parameters are involved. Once we get the parameters of the muffin-tin potential in the Cu/CO system,  it can be straightforwardly used to simulate the case of other 2D lattice.

For convenience, in the following, we will first use the $U_{G_0}$ model to fit the experimental data of the molecule graphene  in Cu(111)/CO system.  Then, with Eq. \ref{ug}, we estimate the parameters of the muffin-tin potential, which is applicable to other 2D lattice.  Finally, using the muffin-tin potential, we give an interesting example in next section.

\subsection{Numerical Simulation}

\begin{figure}
\centering
\includegraphics[width=8.5cm]{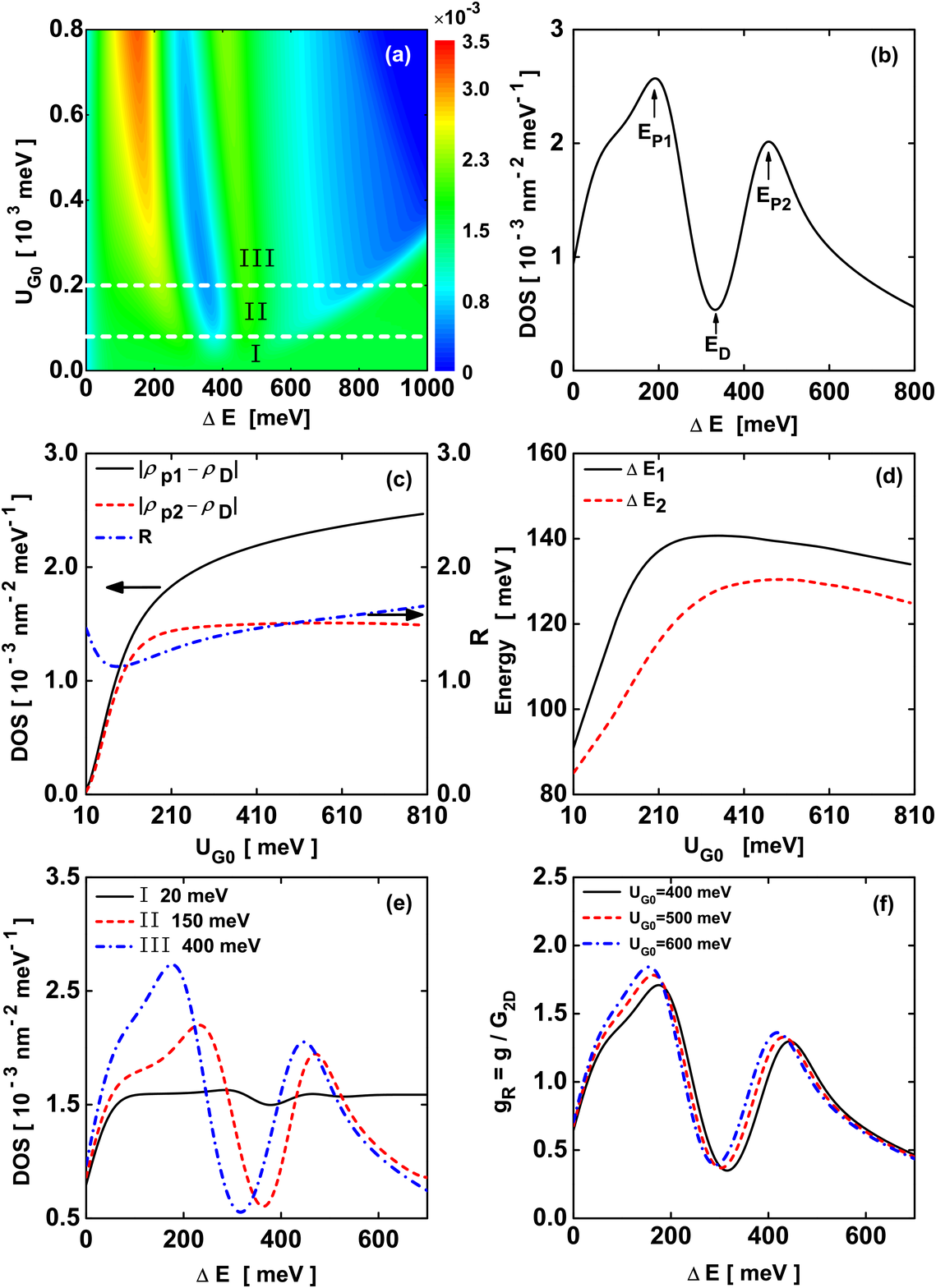}
\caption{(Color online)(a) The DOS of molecule graphene as a function of $U_{G_0}$. I, II, III are three regions of $U_{G_0}$ in which the DOS shape is quantitatively different. (b) An example of the DOS with $U_{G_0}=300$ meV. (c) The respective DOS amplitude of the first and second DOS peaks relative to the DOS minimum and their ratio. (d) Energy positions of the first and second DOS peaks relative to the energy of DOS minimum. (e) Three DOS examples of $U_{G_0}=100, 200, 400 meV$, respectively. (f) The DOS of $U_{G_0}=400, 500, 600 meV$, scaled by the DOS of 2DEG.
}
\label{dos}
\end{figure}

Let us first discuss the experimental results of the molecule graphene in Cu(111)/CO system. In experiment, the molecule graphene is identified by measuring the graphene like DOS, as well as the honeycomb LDOS pattern, with STM\cite{mg2012,lin2014}.
 Thus, in our calculation,  we  only focus on the simulation about the DOS with the effective model, while  the band structure of the molecule graphene has been discussed in detail in former literatures\cite{louie2016,njp2012}. Without the absorbed molecules, the Cu surface states can be approximately viewed as a 2DEG with a parabolic dispersion, the DOS of which is about $\rho_{Cu} = 1.6 \times 10^{-3}$ $\textrm{meV}^{-1}\textrm{nm}^{-2}$. Note that the DOS of pure Cu surface states is used to scale the DOS of the molecule graphene in experiment\cite{mg2012}. The detail DOS data of the molecule graphene is given in Ref. \citenum{mg2012}   with molecule lattice constant $a_0 =1.92$ nm (see the Fig. 1 of the supplementary material of Ref. \citenum{mg2012}). The DOS shape of the molecule graphene is similar as that of graphene, where the DOS minimum is near the Dirac point and two DOS peaks are near the two  Van Hove singularities below and above the Dirac point, respectively.  Beyond the common features, the observed DOS of molecule graphene has some quantitatively characteristics: \textbf{(1)} \emph{the amplitude of the DOS peaks},  $\rho_{P1} - \rho_{D} \approx 1.1 \times \rho_{Cu}$, where $\rho_{P1}$ ($\rho_{P2}$) is the DOS value of the lower (upper) DOS peak relative to the Dirac point. $\rho_D$ is  the minimum of the DOS near the Dirac point; \textbf{(2)} \emph{the two DOS peaks are asymmetry}. The ratio $R= |\rho_{P1} - \rho_{D}| / |\rho_{P2}-\rho_{D}|$ is about $1.9$; \textbf{(3)} \emph{the relative energy positions of the two peaks},  $\Delta E_1 = |E_D - E_{P1}| \approx 110$ meV and $\Delta E_2 = |E_{P2}-E_{D}| \approx 85$ meV, where $E_{P1}$, $E_{P2}$, $E_D$ are the energy of the lower DOS peak , the upper DOS peak, the DOS minimum, respectively.  Then, based on  the effective model of molecule graphene, we would like to see in which parameter region the calculated DOS can reproduce the experimental results.


For the DOS  calculation,  the life time broadening of the energy level, resulting from the scattering between the surface electrons and bulk states,  is a crucial quantity  because that  it determines the magnitude of DOS value.  In experiment, it is estimated that the energy broadening is about $40$ meV  in Cu(111)/CO system\cite{mg2012}.

Now, we discuss the numerical results of the $U_{G_0}$ model, from which we want to estimate the reasonable value of $U_{G_0}$ by comparing with the experimental DOS data. In Fig. \ref{dos} (a),   we  plot the DOS of the molecule graphene as a function of $U_{G_0}$ ($a_0=1.92$ nm), where $\Delta E$ is the energy relative to the band bottom.
 A typical DOS is shown in Fig.\ref{dos} (b), which is like that of graphene.
  In order to do  quantitative comparisons, we plot the amplitude of the two DOS peaks, as well as their ratio $R=\frac{|\rho_{P1} -\rho_D|} { |\rho_{P2} - \rho_D|}$,  in Fig. \ref{dos} (c) as a function of $U_{G_0}$. Meanwhile, the relative energy position of the two DOS peaks are also given in Fig. \ref{dos} (d).   According to these data, we can roughly identify three distinct $U_{G_0}$ regions as indicated in Fig. \ref{dos} (a), and we plot the typical DOS curves in the three regions in Fig. \ref{dos} (e).   In small $U_{G_0}$ region (Region I), the amplitude of the DOS peaks are tiny, which is qualitatively inconsistent with the experiment [the black solid line in Fig. \ref{dos} (e)]. In moderate $U_{G_0}$ region (Region II), the amplitude of the two DOS peaks becomes larger and is comparable to the experimental value, but the two DOS peaks are nearly symmetrical [the red dashed line in Fig. \ref{dos} (e)]. The asymmetry ration R is around 1, which is also qualitatively different from the experimental observation. Only if the $U_{G_0}$ is large enough, we can get a  reasonable DOS shape, which is  qualitatively in agreement with the experimental observation.
  However,  from the DOS data above, we can not get a satisfied quantitative fitting with this effective model, which can exactly reproduce the experimental data.
  For example, if the amplitude of first DOS peak $|\rho_{P1} - \rho_D|$ is exactly equal to the experimental value, the $U_{G_0}$ should be in the moderate $U_{G_0}$ region and give a symmetrical DOS shape. Meanwhile, in the large $U_{G_0}$ region, the  relative position of the two peaks are about 20 meV larger than the experimental value (from 20 meV to 30 meV depending on $U_{G_0}$ value).

  Considering all the numerical results above, we suggest that, for the molecule graphene in Cu/CO system, the acceptable region of $U_{G_0}$ is between 400 meV to 600 meV. Only if $U_{G_0}$ is in this region, we can get a reasonable DOS shape, while the peak amplitude, peak position and asymmetry ratio is comparable to the experimental value.  Two important issues should be made clear further. One is that, in former studies,   the value $U_{G_0}$ is estimated by the energy difference between the M point and Dirac point in the BZ, and it is argued that $U_{G_0}$ is about 100 meV\cite{mg2012,polini2013}.
However, when the life time broadening is considered in the DOS calculation, it is obvious that the DOS peaks are not at the M point and the DOS minimum is also shifted from the Dirac point. That is why the $U_{G_0}$ value suggested here is larger than the former one. We plot the DOS with suggested $U_{G_0}$ values in Fig. \ref{dos} (f). The other is that even in this $U_{G_0}$ region, there is still a  difference about 20 meV between the experimental and theoretical values of the  position of the  DOS peaks. We argue that this discrepancy may result from the assumption about the life time broadening, where it is considered as a constant in our calculation.     Actually, the life time broadening of the metal surface states has been  intensively studied, and normally it is a function of energy\cite{lifetime2002}. To get a better effective model is still an open question and need further efforts in future.

\begin{figure}
\centering
\includegraphics[width=8.7cm]{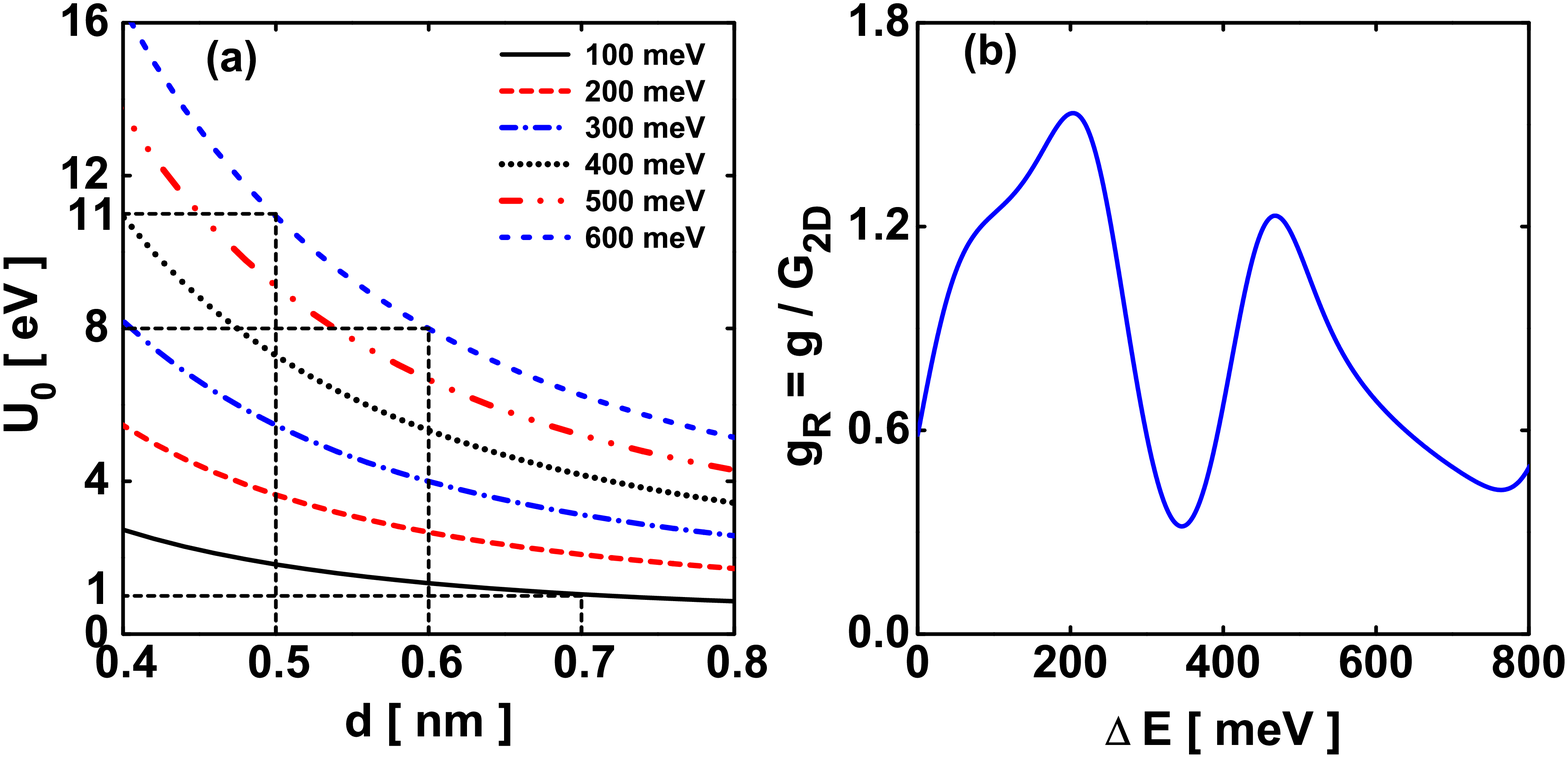}
\caption{(Color online) (a) Molecule potential $U_0$ as a function of the potential diameter $d$ for $U_{G_0}=100,200,300,400,500,600 meV$ respectively. (b) The DOS of the muffin-tin model with our estimated parameters $U_0=8.0 eV$, $d=0.6 nm$, $a=1.92 nm$, scaled by DOS of 2DEG.
}
\label{u0}
\end{figure}

We then estimate the parameters of the muffin-tin potential with Eq. \ref{ug}.  Except the lattice constant $a_0$, which is controlled in experiment, the muffin-tin potential model has two parameters: potential value $U_0$ and diameter $d$. Given $U_{G_0}$, the Eq. \ref{ug} indicates that $U_0$ is a function of $d$.  In Fig. \ref{u0} (a), we plot $U_0$ as a function of $d$ for different $U_{G_0}$ values. According to the charge difference got by DFT calculation, the potential diameter $d$ of the CO molecule should be in the region between 0.5 nm to 0.7 nm\cite{DFT2014}.
 As indicated in Fig. \ref{u0} (a), we  can see that the  $U_0$ should be in the region between 4 eV and 10 eV. Note that the $U_0$ is larger than 1 eV even if  $U_{G_0}=100$ meV. Thus, it is safely to say that the $U_0$ of muffin-tin potential is of several eV for the Cu/CO system, which is the central result for this section.  Here, we give some further evidence from the related quantum corral system to support our estimation above\cite{corralrmp,corral1996, corral2004}.
 In the well-known quantum corral system, adatoms (e.g Fe atoms) are absorbed on Cu(111) surface to form a corral, which confines the surface electrons and gives rise to a standing wave pattern in the corral.  Similar as the molecule graphene, each adatom  applies a local potential on the surface electrons. By fitting the experiment, it is suggested that the potential value of the adatom is about several eV\cite{corral1996,corral2004}. Together with our calculation, we believe that our estimation about the $U_0$ is reasonable. Note that the potential value $U_0$ are different for different molecules. Our estimation is only applicable for the Cu/CO system. In the  following, we assume using Cu/CO system by default, and use $U_0=8$ eV and $d=0.6$ nm for further study, the DOS of which is plotted in Fig. \ref{u0} (b).

\section{An example: Artificial Kagome lattice }

In last section, we discuss the effective model of molecule graphene.  For the Cu/CO system, we estimate the  reasonable value for  $U_0$  and $d$, which is applicable to all the 2D lattice. In this part, we would like to give an interesting example which is beyond the scope of graphene physics.  We design a 2D molecule lattice, and by simulation with muffin-tin model, we show that this 2D lattice can mimic the band structure of Kagome lattice. As well known, the Kagome lattice is of highly interests in condensed matter physics because of its novel correlated ground state\cite{kagome1992,spinliquid2011,coldatomkagome,flatband2014}. So far, in solid state electron system, a clean Kagome lattice, as simple as graphene, has not been found.  Our scheme may offer a novel platform to study the physics of Kagome lattice.

\begin{figure}
\centering
\includegraphics[width=8cm]{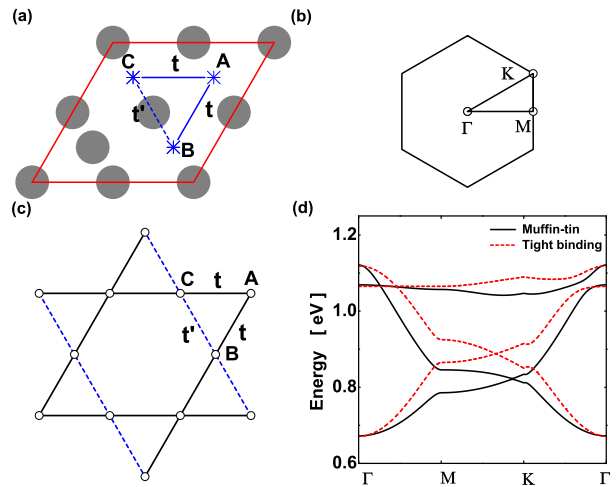}
\caption{(Color online) (a) A muffin-tin potential designed to transform the hexagonal lattice to Kagome lattice. The potential is $U_0>0$ inside the gray disks and zero elsewhere. Red lines illustrate the unit cell of the molecule lattice. Blue stars indicate the electron sites in the thin film of the surface. (b) The corresponding Brillouin zone. (c) The corresponding Kagome lattice of the given muffin-tin potential. (d) Black solid lines are energy bands for the designed muffin-tin potential with estimated parameters $U_0=8.0 eV$, $d=0.6 nm$, $a=1.3 nm$. Red dashed lines are bands calculated by tight binding to fit muffin-tin bands. The parameters are $t=78.5 meV, t'=50 meV$. On-site energy $\epsilon_A=80 meV, \epsilon_B=\epsilon_C=40 meV$.
}
\label{kagome}
\end{figure}

The molecule arrangement we designed is shown in Fig. \ref{kagome} (a), where each gray disk represents a molecule and the unit cell is indicated by the red line. The basic idea is simple.
First, we arrange the molecules to form a hexagonal lattice.
If we assume that  each lattice site of the surface electron is positioned at the inner of diamond formed by four adjacent molecules, we then get a same hexagonal lattice for the surface electrons. It should be noted that if one quarter of the lattice sites is deleted, the hexagonal lattice can be transformed into the Kagome lattice. Thus, we use an additional molecule to ``kill" one lattice site, as shown in Fig. \ref{kagome} (a). There are three sites in one unit cell of the Kagome lattice, which is marked by the blue stars in Fig. \ref{kagome} (a) (A, B, C sites).  Meanwhile, the nearest neighbourhood  hopping are denoted as $t$ and $t'$,  where the hopping between B and C sites ($t'$) is different from that between A and B ($t$). The corresponding Kagome lattice is given in Fig. \ref{kagome} (c).

 \begin{figure}
\centering
\includegraphics[width=6cm]{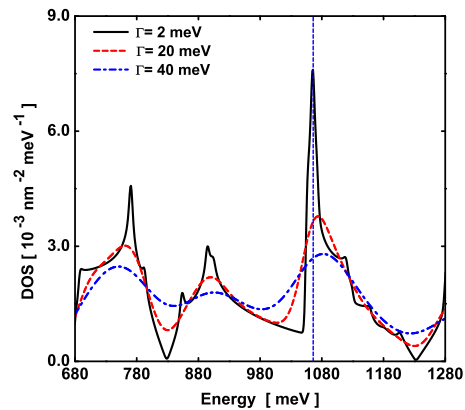}
\caption{(Color online) The DOS of the designed muffin-tin potential with estimated parameters $U_0=8.0 eV$, $d=0.6 nm$, $a=1.3 nm$. Black solid line is DOS with life time broadening $\Gamma=2 meV$, and red dashed line $\Gamma=20 meV$, blue dashed dotted line $\Gamma=40 meV$. Blue dotted line is the fermi level of the DOS with $\Gamma=40 meV$.
}
\label{kDOS}
\end{figure}

We then use the muffin-tin potential with the parameters of Cu/CO system estimated above to simulate the artificial Kagome lattice system.  The lowest three bands are plotted in Fig. \ref{kagome} (d) (the black solid line).  We try to use a  tight binding model to fit the calculated band, and the results are plotted in Fig. \ref{kagome} (d) (the red dotted line).  The tight binding Hamiltonian is
\begin{equation}
H=\sum_{i, \eta\in{ABC}, \sigma} \epsilon_{\eta} c^+_{i\eta\sigma}c_{i\eta \sigma} + H_t
\end{equation}
where $i$ is the index of unit cell, $\eta$ indicates the three different sites in one unit cell and $\sigma$ denotes the electron spin.   $H_t$ is the hopping terms
\begin{equation}
H_t =  -t \sum_{\langle i,j \rangle} ( c^+_{iA\sigma}c_{jB \sigma} +  c^+_{iA\sigma}c_{jC \sigma} ) -t' \sum_{\langle i,j \rangle} c^+_{iB\sigma}c_{jC \sigma} + h.c.
\end{equation}
where $\langle i,j \rangle$ denotes the nearest neighbourhood hopping.
 In spite of our best efforts, we only get a  qualitatively fitting about the energy bands, which is not as good as the case of molecule graphene.  The parameters are $t=78.5$ meV,  $t'=50$ meV, $\epsilon_A=80$ meV and $\epsilon_B=\epsilon_C=40$ meV. The numerical simulation demonstrates that the molecule structure we designed can be roughly viewed as a Kagome lattice with  non-equal nearest neighbourhood hopping. The corresponding DOS is given in Fig. \ref{kDOS} with different life time broadening. For Kagome lattice, one of the three bands is very flat, and thus induce a giant DOS peak (see in Fig. \ref{kDOS}), which may induce interesting correlated electronic states in experiment. Meanwhile, the Fermi level is controllable by adjusting the molecule distance $a_0$.

\begin{figure}
\centering
\includegraphics[width=6cm]{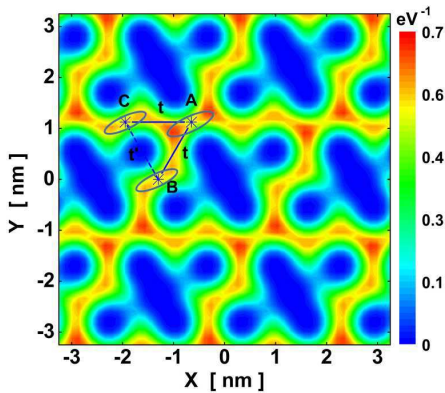}
\caption{(Color online) The LDOS of the designed muffin-tin potential with estimated parameters $U_0=8.0 eV$, $d=0.6 nm$, $a=1.3 nm$ at the energy level $E=850 meV$.
}
\label{kLDOS}
\end{figure}

 The main reason for the discrepancy between the tight binding bands and the ones got by plane wave method is because that each site of the surface electrons here has  inner electron structure. This can be seen from the LDOS of this Kagome lattice, as shown in Fig. \ref{kLDOS}. In each site, e.g. the B site, the electron distribution is not uniform, and seems to be composed of two small sites. However, we try several possible tight bind models and do not get a better approximation.  Anyway, ignoring the detailed inner electronic structure in each site, the LDOS can be viewed as a Kagome lattice, which is consistent with the energy band calculation and can be described by the tight binding model above. Since in our simulation we use the realistic parameters of Cu/CO system, we believe that the Kagome lattice structure proposed here can be easily tested in experiment.

\section{conclusion}
In summary, the  purpose of this work is to point out that the technique used to build molecule graphene in recent experiments offers a new opportunity to explore the novel physics in any 2D lattice beyond the scope of graphene.
To demonstrate this idea, we first analyze the effective models of the molecule graphene. We discuss the relation and distinction between the two potential models used to simulate the molecule graphene. Then, by fitting the DOS data, we estimate the parameters for each potential model in the Cu/CO system. We point out that, for the muffin-tin potential, potential value  $U_0$ is of  the order of several eV and potential diameter $d$ is between 0.5 and 0.7 nm, which can be used to simulate any other 2D molecule lattice in the Cu/CO system.  Finally, we give an interesting example. We design a molecule lattice, and by numerical simulation we show that this molecule arrangement can give a Kagome like band structure.
Note that in experiment there are two ways to build a molecular lattice on metal surface. One is by the STM tips, as discussed above. The other is to use the self-assembly process, where organic porous networks are used to confine the surface electrons\cite{porous2009,wang2013}. However,  it is still a big challenge to build a complex 2D lattice of surface electron by the self-assembly technique, though it is useful to make a molecule network in a large area.

The molecule graphene represents a new kind of artificial 2D electron lattice system, which has its own advantages compared with other artificial 2D lattice systems, e.g. the cold atom, nano-patterned 2DEG in semiconductor, and photonic crystal. It of course will not only limited to the graphene system, but also can be used to study any novel 2D lattice.
From the success of molecule graphene, we believe that the technique is now ready. But it should be noted that the molecule graphene is a quantum antidot system. To build the desired 2D lattice, we need a map from the molecule arrangement to the 2D surface electron lattice we want. That is what we do in this work.   Our scheme here give a theoretical proposal to get a Kagome like 2D lattice. And   the estimation about the model parameters of the Cu/CO system  also gives a basis for future numerical simulation.   We hope that our work can stimulate more research interest into this new artificial 2D lattice system, and believe that it will lead to new experimental progress in near future.

\begin{acknowledgements}
 This work is supported by the National Science Foundation of China (Grants No. 11274129, No. 11534001).
\end{acknowledgements}

\bibliography{effectivemodel}

\end{document}